# Binding Isotherms and Cooperative Effects for Metal-DNA Complexes.

Eteri S. Gelagutashvili

E.L. Andronikashvili Institute of Physics, 6 Tamarashvili str.,

Tbilisi, 0177, Georgia

E.mail: gel@iphac.ge

gelaguta@yahoo.com

## Abstract

The stoichiometric binding constants of Nickel(II), Cobalt(II), Manganese(II), Silver(I), Zinc(II) ions with DNA, from *Spirulina platensis* were determined from their binding isotherms by equilibrium dialysis and atomic absorption spectroscopy. It was shown, that the nature of these ions interaction with DNA, from *S .platensis* is different. For Cobalt(II), Zinc(II) ions were observed cooperative effects and existence of two different types of the binding sites. Nickel(II)_, Silver(I) -DNA complexes shows independent and identical binding sites and Manganese(II)_ negative cooperative interaction.

The logarithm of binding constants for Cobalt (II)_, Nickel (II)_, Manganese (II)_, Zinc (II)_, Silver (I) - DNA, from *S. platensis* in 3 mM Na(I) are 5.11; 5.18; 4.77; 5.05; 5.42; respectively.

The linear correlation of *logarithm of binding constants* (for complexes of metal-DNA from *S. platensis*) and the covalent index of Pauling are observed.

Interaction of metal ions with DNA is an important fundamental issue in life sciences related to the replication, transcription and translation of DNA, mutation of genes, action mechanisms of some drugs etc.

In *Cyanobacterium Synechococcus sp*. PCC 7942, *SmtB*, functioning as a sensor to heavy-metal ions in the dimmer form represses transcription of *smtA* gene encoding metallothionein-like protein [1]. There are two recognition DNA sequences in the operator/promoter region of smtA [2]. Site directed mutagenesis experiments demonstrated that the specific recognition site for ManR was TATGAAAGAATATGAGAA composed of two direct repeats of the consensus sequence (TA) ATGA(GA)A(AG). This is a novel regulatory DNA motif in cyanobacteria, indicating that the expression of mntH is regulated by two-component Mn(II)-Sensing System containing *ManR* in *Anabaena sp.*PCC 7120. This specific pathway of regulating mntH expression has been found only in cyanobacteria [3]. The mechanism of metal ion selectivity by members of the *SmtB/ArsR* family of bacterial metal sensing transcriptional repressors and the mechanism of negative allosteric regulation of DNA binding is poorly understood [4].

One of the oldest living plants on the planet *Spirulina platensis* is a filamentous cyanobacterium that is important for biothecnology due its high nutritional value. Although *S. platensis,* due to its ubiquitous occurrence in nature, is studied extensively the nature of metal ions interaction with them and their components is not known.

In this paper the energetics of binding of Ag(I), Co(II), Ni(II), Mn(II) and Zn(II) ions to DNA isolated from blue-green algae *S. platensis* was determined from their binding isotherms by equilibrium dialysis and Atomic-absorption spectroscopy.

## Materials and Methods

Chloride salts of Co(II), Ni(II), Mn(II), Zn(II), and Na(I) and nitrate salts of Ag(I) and Na(I) were used as the reagents. All reagents were of analytical grade and prepared in double-distilled water.

The description of technique that was used for DNA isolation from *S. platensis* permitting to obtain sufficient yield of the preparation was compiled according to the method described in article [5] with some modifications. Spirulina biomass (preferably harvested during the logarithmic phase of growth) is centrifuged at 100**g** for 3-5 minutes. 5-10g of cell precipitate is washed by suspending it first in distilled water with following centrifugation and then the same procedure is repeated in buffer (TRIS-EDTA, pH 9.0). To destroy cell envelope suspension of cells in NaCl-EDTA solution (? 35ml of solution per 5 g of cell raw weight) is quickly frozen in the mixture liquid nitrogen-acetone or artificial ice-acetone ($-70^0$?) and then it is unfrozen at $+37^0$? with the same rate. For complete cell destruction lysozyme is added, about 40 g of the enzyme per 5 g of cell dry weight. The mixture is incubated at $+37^0$? for 2 hours. In order to separate protein from DNA, 25% solution of SDS is added to cell lysate up to final concentration of 2% followed by heating in water bath at $60^0$? for 10 minutes. To deproteinize DNA preparation 5? NaCl or 5? $NaClO_4$ is added to the suspension up to final concentration of 1M followed by subsequent treatment with water-saturated phenol and mixture of chloroform-isoamyl ethanol (24:1). The procedure is performed adding equal volume of phenol and shaking the mixture in cold on the mechanical shaker for 30 minutes. Then the mixture is centrifuged at 3000 **g** for 30 minutes. The water phase, containing DNA with RNA impurity, is carefully drained off by a pipette with a dished end. The obtained solution is repeatedly deproteinized by the above-described method. The procedure is repeated until protein interlayer between phenol and DNA-containing upper phase disappears. The upper phase (deproteinized DNA) is carefully drawn off, to precipitate DNA fibers this solution is carefully poured into double volume of 90% ethanol and stored for 20-30 min in cold for formation of DNA precipitate ("medusa"). In this case pigments are not adsorbed on the DNA surface. The "medusa" is carefully removed to a dry glass by a glass stick with dished end, then we let ethanol to drain, quickly wash DNA with NaCl-acetate solution and dissolve it in 5-8 ml of the same solution. To remove impurities of RNA and polysaccharides DNA is incubated with ribonuclease (50 µg/ml) and ?-amylase (200µg/ml) at $37^0?$ for 30 minutes. After the incubation pronase (100µg/ml) is added for complete removal of ferments and proteins, and then mixture is stored at $37^0?$ for 2 hour more (pronase being previously heated at $60^0?$ for 30 minutes). After the solution is cooled, DNA deproteinization is performed again by shaking it with equal volume of mixture chloroform-isoamyl ethanol (24:1) and centrifugation for 15 minutes. DNA is precipitated from the upper layer by 90% ethanol. "Medusa" is washed in 70% ethanol and then dissolved in small volume of NaCl-citrate solution. To remove RNA impurities 1 ml of acetate-EDTA (pH 7.0) is added to DNA solution, thoroughly mixed and then DNA is re-precipitated by 10 ml of isoamyl ethanol. When doing so we precipitate only DNA, whereas RNA fractures, pigments and polysaccharides remain in the solution. The isolated DNA fibers are washed in 75, 80, 90 and 96% ethanol, dissolved in NaCl-citrate solution with several drops of chloroform added, to be stored for a considerable period of time. DNA preparations were evaluated by spectral indicators, which were in accordance with literature data.

Equilibrium dialysis experiments were performed in a two–chambered Plexiglass apparatus. The chamber capacity was 5ml. The membrane thickness was 30 ? m (Visking). One chamber contained DNA ($10^{-4}$M) and the other –solution of the metal ion under investigation. The initial metal concentration varied within the range $10^{-6}$-$10^{-4}$ M. Samples were analyzed by flame atomic–absorption spectrophotometry (FAAS) ("Beckman") for Mn, Co, Ni, Zn, Ag on 279.5, 240.7, 232, 213.8, 328,1 nm respectively.

<u>Data analysis</u>. Binding constants were determined from the Scatchard plots. The equilibrium binding of metal ions with independent and identical binding sites in DNA molecules (so-called Scatchard plots) can be written as:

$$r/m = K(n-r), \qquad (1)$$

where *k* is stoichiometric binding constant, *n* is the number of metal binding sites per phosphate group of DNA at saturation, *r* is the concentration of bound metal ions, *m* is the concentration of free metal ions.

Modified Scatchard equation in the case of two types of binding is:

$$r/m = 0.5\,[\,B(r) + \sqrt{B^2(r) + 4C(r)}\,], \qquad (2)$$

where $B(r) = k_1 n_1 + k_2 n_2 - (k_1 + k_2)\,r;$

$C(r) = k_1 k_2 r (n_1 + n_2 - r)$. $k_1 k_2$ are mikroconstants and $n_1, n_2$ the number of binding sites for metal ions per phosphate group of DNA.

In the case of negative cooperative binding is:

$$r/m = K e^{-Wr}(n-r), \qquad (3)$$

where $W$ is the constant depending on the repulsion energy between the bound metal ions [6,7].

## Results and discussions

The adsorption isotherms of Zn(II)-, Co(II)-, Ni(II)-, Ag(I)-, Mn(II)-DNA complexes in the Scatchard coordinates at 3mM Na(I) and t=20°C are shown in Fig.1-4.

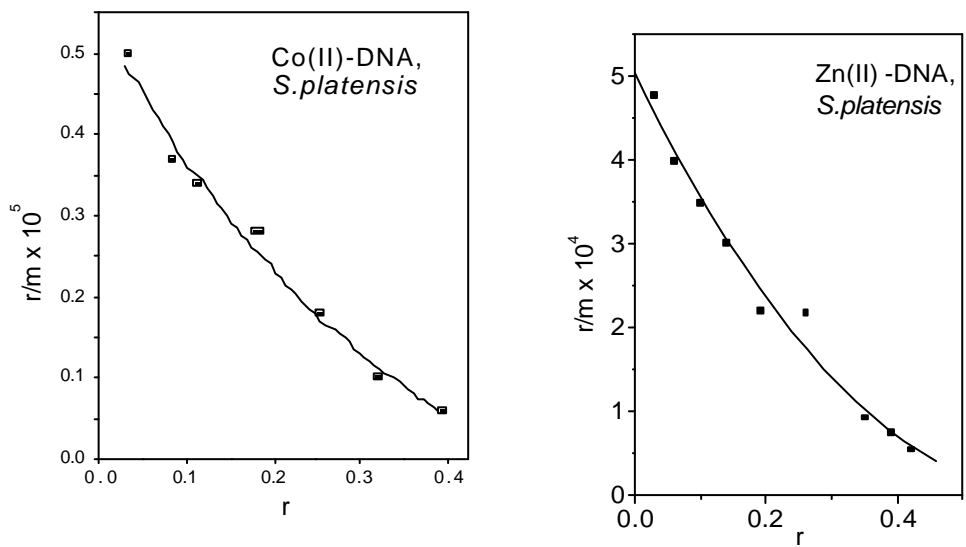

Fig.1. Binding isotherms of Co(II) – and Zn(II)-DNA complexes in
the Scatchard coordinates. *r* is the bound metal ions concentration,
*m* – the concentration of free metal ions. The points
show experimental data and the plots are received using eq. (2)

(in all cases each point represents the average of three independent determinations. Standard deviations were <11% of the means). As it is seen from Fig. 1, the dependence of *r/m* on *r* is nonlinear for Co(II) and Zn(II). Nonlinearity of the binding plot may be caused by several effects that are difficult to distinguish: overlaping of binding sites, cooperative effects and existence of two different types of the binding sites. In fig.1, points are experimental and curves are obtained using eq.2. Using equation (2) micro constants $k_1, k_2$ and corresponding numbers of binding sites $n_1, n_2$ for Co(II) and Zn(II) were determined (table 1). Similar type of interaction was observed for Pb(II)-DNA complexes[8].

As it is seen from table 1, $k_1 > k_2$ for Zn(II), Co(II), Pb(II) ions, i.e. the association of these ions with DNA can be described by the model with two patterns of binding, one of them corresponding to the strong binding, the other corresponding to the weak one. Two types of binding is also observed in case of Cu(II)-DNA complexes, from *S. platensis* [9] (LogK $_{Cu}$=5.19). In case of Co(II) ions linear dependence between *r/m vs r* was observed in area 0<r <0.06. It is in good agreement with literature data, where interaction between Na-DNA and Co(II) cations has been investigated by $^{23}$Na NMR relaxation [10]. A linear dependence of $?R_{obs}$ on r is observed in concentration range *0< r <0.08.*

**Table 1**. Binding parameters for Zn(II)-, Pb(II)- and Co(II)-DNA complexes at 3 mM ionic strengths, t = 20°C.

| | Co(II) - DNA | Zn(II) - DNA | Pb(II)- DNA[8] |
|---|---|---|---|
| Micro constant $k_1$ ´$10^4$, ? $^{-1}$ | 73.3 | 18.0 | 24.1 |
| Micro constant $k_2$ ´$10^4$, ? $^{-1}$ | 7.1 | 10.31 | 13.3 |
| Number of binding sites $n_1$ | 0.04 | 0.05 | 0.07 |
| Number of binding sites $n_2$ | 0.43 | 0.41 | 0.42 |
| Stoichiometric binding constant $K$ ´$10^4$, ? $^{-1}$ | 12.8 | 11.15 | 14.38 |
| *log K* | 5.11 | 5.05 | 5.16 |
| Gibbs free energy -$DG°$ kcal/mol | 6.95 | 6.87 | 7.02 |
| $c^2$ distribution | 0.14 | 0.05 | 0.008 |
| Correlation coefficient R | 0.96 | 0.98 | 0.98 |

It is seen from fig.2 that at low ratio of occupied sites corresponding up to one bounded nickel per 19 DNA phosphorus, a positive slope was observed indicating positive cooperativity of Ni(II) binding to DNA. Analogous type of interaction was obtained also for Cd(II)-DNA complexes[9] (Log$K_{Cd}$=5.16). When the proportion of the bounded nickel ions per DNA phosphorus fell within the range between 1:19 and 1:2, an approximate linear relationship with negative slope was observed. The best fit slope for the descending portion of the binding curve revealed an affinity constant K= 15.1x $10^4$ $M^{-1}$ for Ni(II). ( table 2).

Table 2. Parameters of Mn(II), Ag(II), Ni(II) and Cr(III) ions binding to DNA at 3 mM ionic strengths, t=$20^0$C.

|  | Mn(II) | Ag(I) | Cr(III)[16] | Ni(II) |
|---|---|---|---|---|
| Stoichiometric binding Constant $Kx10^4$, $M^{-1}$ | 5.89 | 26.7 | 9.1 | 15.1 |
| Log K | 4.77 | 5.42 | 4.96 | 5.18 |
| Gibbs free energy -$DG^0$ kcal/mol | 6.49 | 7.38 | 6.75 | 7.04 |
| Repulsion energy W | 10.2 | - | - | - |
| Number of binding sites n | 0.45 | 0.49 | 0.54 | 0.5 |
| Correlation coefficient R | 0.95 | 0.97 | 0.98 | 0.97 |

The type of the dependence of *r/m* vs *r*, means that there is negative cooperative interaction between bound Mn(II) ions. In. fig.3 the points are experimental data and the plot is received using eq.3. All these parameters are presented in table 2. Every pair of nucleotides is associated with manganese at saturation. Value of *W* is in good agreement with literature data [6,7] and with our previous data for Cu(II)-, Ni(II)-, Co(II)- and Zn(II)-nucleosome complexes[11,12]. At saturation, one metal ion corresponds to 2 bases for all metal ions. The same value of *n* was obtained in [13,14,15], where the interaction of Pb(II), Ni(II), Cu(II) and Zn(II) ions with DNA from calf thymus was studied.

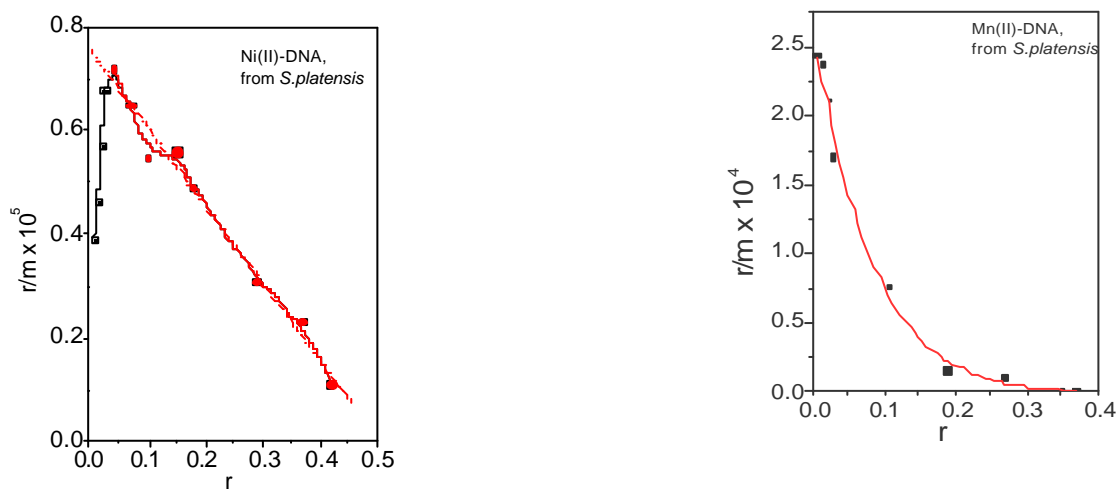

Fig.2 Binding isotherms of Ni(II)-DNA complexes. The parameters are the same, as in fig.1. The points show experimental data and the plot is received using eq.(1).

Fig.3 Binding isotherm of Mn(II)-DNA complexes. The parameters are the same as in fig.1. The points are experimental data and the plot is received using eq.3.

Analysis of the plot for Ag(I)-DNA complexes (fig.4) shows independent and identical binding sites. Interaction between Cr(III) ion and DNA is of the similar type[16]. The best-fit slope for Ag(I) and Cr(III) revealed affinity constants K = *26.7x10$^4$ M$^1$* and *K = 9.1x10$^4$ M$^1$* respectively (table 2).

Metal ions may be classified according to hard and soft acid/base schemes. Part of them is likely to bind predominantly relatively „hard'' phosphate groups. On the other hand, metal ions may bind preferentially to the „softer'' nitrogen sites in the bases. It is known, that Co(II), Zn(II),

Mn(II), Pb(II) and Ni(II) ions represent group of borderline ions, which display ambivalence to donors of O, N - and S-type. Specific binding sites in DNA for borderline metal ions are phosphorous groups and nucleotide bases. The study of Co(II) binding to DNA using high resolution X-ray diffraction [17] showed, that they bind exclusively to N7 of guanines with direct coordination at different sequence locations. Interaction of divalent transition metal ions (such as Ni(II), Mn(II), Co(II), etc.) with AT-DNA bases is nonspecific and predominantly electrostatic[18]. Guanine N7 is the preferred binding site for transition metals [19]. Particularly, Mn(II) and Zn(II) ions bind selectively to G4-N7 of the DNA oligonucleotide d[CGCGAATTCGCG] [20]. In GC DNA, Ni(II) binds strongly and specifically to the N7 atom of guanine in the major groove[18]. Published data on Ag(I) and Cr(III) coordination sites are contradictory. *Ab initio,* the calculations were performed to characterize structure and energetics of Ag(I) ion complexes with noncomplimentary DNA base pairs (cytosine-adenine) and *an aqua* ligands of silver coordination sphere within the Hartry-Fock approximation. The calculation showed that in all structures modified by Ag(I), metal-base interaction was dominant compared with metal-water interaction, whereas the inter ligand repulsion was not significant[20]. Fourier transform infrared spectroscopy and capillary electrophoresis were used to analyze the Ag(I) binding mode, the binding constant, and the polynucleotides' structural changes in the Ag-DNA complexes. The spectroscopic results showed that at type 1 complex formation with DNA, Ag(I) binds to guanine N7 at low cation concentration ($r = 1/80$) and to adenine N7 site at higher concentrations ($r = 1/20$ to $1/10$), but not to the backbone phosphate group. At $r = 1/2$, type II complexes form with DNA in which Ag(I) binds to the G-C and A-T base pairs[21]. Scatchard analysis of capillary electrophoresis data showed two binding sites for Ag-DNA complexes with $K_1 = 8.3 \times 10^4$ M$^{-1}$ for guanine and $K_2 = 1.5 \times 10^4$ M$^{-1}$ for adenine bases. The interaction between silver ion and DNA has been followed by submarine gel electrophoresis [22]. It was shown, that the mobility of the bands decreased as the concentration of Ag(I) was increased, indicating the occurrence of increased covalent binding of the metal ion with DNA.

Earlier studies of DNA binding with chromium complexes suggested that guanines were the preferred binding sites for chromium[23], more recent studies demonstrated that chromium binding to DNA were not base selective and that the primary chromium binding sites were the phospate groups[24]. Recent data [25] show, that Pb(II) forms hemidirect coordination compounds both in complexes with water and in complexes with O- and N- atoms of ligands, which, in their turn, are reactive centers in DNA for other metal ions [26]. In [8], it was shown that with increase of GC content in DNA stoichiometric binding constant for Pb(II) increased. The binding constants increase with increase of GC content due to preferred association of Pb(II) ions with GC pairs. GC content in DNA of *S.platensis* exceeds that of calf thymus. Hence, it may be expected that such effective binding at increase of GC content is a result of the raise of probability of N7-transition metal ion complex formation.

In general, binding between cations and poly-ions (DNA) may be inner-sphere (direct interaction with active groups of DNA), outer-sphere (territorial – interaction of metal ions via the bridging action of water molecules) and "atmospheric" (through long-range electrostatic interaction only) binding [7].

Table 1 and 2 summarize the obtained data as standard Gibbs energy. These results show that the value of $DG°$ is of the same order as the energy of hydrogen bonding i.e. Zn(II), Cr(III), Co(II), Ni(II), Mn(II) and Pb(II) ions form mainly outer sphere complexes with DNA and $DG°$ exceeds the energy of hydrogen bonding for Ag(I). 0n the basis of $pK$, metals are arranged in the descending order as follows:

**Ag(I)>Cu(II) >Pb(II), Cd(II) >Ni(II)> Co(II)> Zn(II) > Cr(III)>Mn(II)**.

It can be assumed that the specific role of metal ions may be attributed to the individual balance of interaction between cations and the active centers of DNA.

In [27, 28] correlation between the covalent index $(X_m^2 r)$ and phase transition midpoints are observed. $X_m^2 r$ is an indicator that compares energy of valence orbital with energy of ionic bond [27, 28]. Analogous dependence between stoichiometric binding constants and $X_m^2 r$ was obtained in our case.

The linear correlation of $–logK = pK$ (for complexes of metal-DNA from *S. platensis*) and the covalen?? index $X_m^2 r$ are presented in Fig.5. (Correlation coefficient R -0.87; P 0.002; SD 0.36).

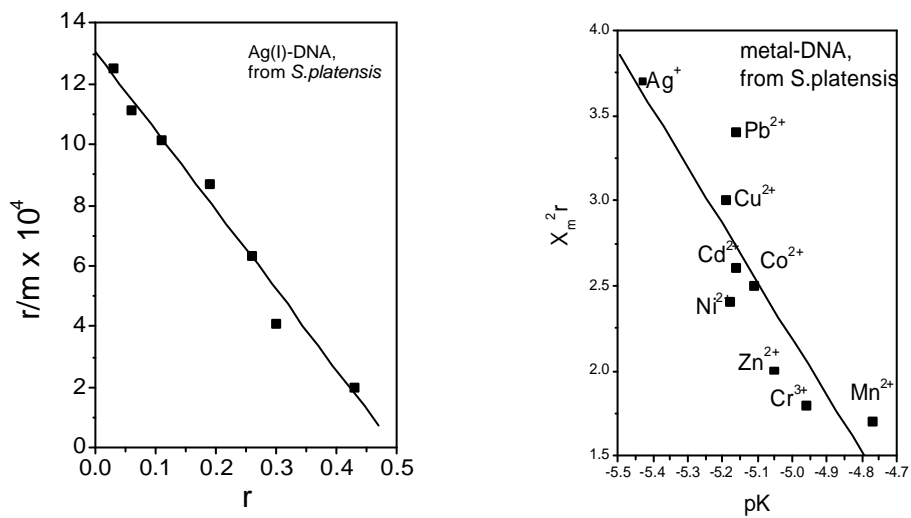

Fig.4. Binding isotherm of Ag(I)-DNA complexes. The parameters are the same as in fig.1. The points are experimental data and the plot is received using eq. 1.

Fig.5. The Dependence the covalen?? index $X_m^2 r$ vs $-logK = pK$ (for complexes of metal-DNA from *S. platensis*).

**Acknowledgments**

The author thank Dr. A. Belokobilsky for extraction of DNA from *Spirulina platensis.*